\documentclass[fleqn,usenatbib]{mnras}

\usepackage{newtxtext,newtxmath}

\usepackage[T1]{fontenc}
\usepackage{ae,aecompl}
\usepackage{rotating}
\usepackage{color}

\usepackage{graphicx}	
\usepackage{amsmath}	
\usepackage[utf8]{inputenc}
\usepackage[dvipsnames]{xcolor}
\definecolor{royalfuchsia}{rgb}{0.79, 0.17, 0.57}

\title{Detection of 125.5-day Optical Periodic Modulation of the Neutron Star M51 ULX-8}

\author[S. Allak]{S. Allak,$^{1,2}$\thanks{E-mail:0417allaksinan@gmail.com}
\\
$^1$Department of Physics, University of Çanakkale Onsekiz Mart, 17100, Çanakkale, Türkiye \\
$^2$Space Science and Solar Energy Research and Application Center (UZAYMER), University of Çukurova, 01330, Adana, Türkiye\\
}
\date{Accepted XXX. Received YYY; in original form ZZZ}

\pubyear{2022}

\begin{document}
\label{firstpage}
\pagerange{\pageref{firstpage}--\pageref{lastpage}}
\maketitle

\begin{abstract}

Studying Ultraluminous X-ray sources (ULXs) in the optical wavelengths provides important clues about the accretion mechanisms and the evolutionary processes of X-ray binary systems. In this study, three (C1, C2, and C3) possible optical counterparts were identified for well-known neutron star (NS) candidate M51 ULX-8 through advanced astrometry based on the {\it Chandra} and Hubble Space Telescope ({\it HST}) observations, as well as the {\it GAIA} optical source catalogue. Optical periodic modulation of 125.5 days with an amplitude of 0.14 magnitude was determined for C3 which has evidence to represent the optical nature of ULX-8 using one-year (2016-2017) 34 {\it HST} ACS (Advanced Camera for Surveys)/WFC (Wide Field Camera) observations. Moreover, surprisingly, the observed optical fluxes of C3 exhibit a bi-modal distribution. This could mean that there is a possible correlation between the optical and the X-ray flux variabilities of the ULX-8. The possible scenarios which are frequently mentioned in the literature proposed for the nature of optical emission and optical super-orbital period. The most probable scenario is that the optical emission could have originated from the accretion disk of the ULX-8.

\end{abstract}


\section{Introduction}

Ultraluminous X-ray sources (ULXs) are non-nuclear point-like extragalactic sources with an isotropic X-ray luminosity (L$_{X}$ > 10$^{39}$ erg s$^{-1}$) are exceeding the Eddington limit for a typical 10\(M_\odot\) stellar-mass black hole (sMBH) \citep{2017ARA&A..55..303K}. Many studies were presented evidence of intermediate mass black holes (IMBHs, 10$^{2}$- 10$^{5}$ M$\odot$) with accreting at sub-Eddington rates for the compact nature of ULX systems \citep{1999ApJ...519...89C,2004ApJ...614L.117M,2007Ap&SS.311..203R,2012MNRAS.423.1154S,2013MNRAS.436.3262C,2013MNRAS.436.3782D}.

Thanks to the new-generation of X-ray observatories (e.g. {\it XMM-Newton}, {\it Chandra}, {\it NuSTAR}), the nature of ULXs has been further studied and a new perspective has been developed including accreting scenarios at super-Eddington limit for a sMBHs \citep{2014ApJ...793...21W,2015MNRAS.454.3134M,2018ApJ...857L...3W}. The discovery of a NS ULX (or pulsar ULX, PULX) in M82 by \cite{2014Natur.514..202B} led us to change our perspective on the nature of ULXs. It is also exciting evidence that highly super-Eddington accretion can occur for sMBHs. Thrilling discoveries of PULXs continued in the followed years \citep{2016ApJ...831L..14F,2017Sci...355..817I,2017MNRAS.466L..48I,2018MNRAS.476L..45C,2019MNRAS.488L..35S,2020ApJ...895...60R}. Moreover, a cyclotron resonance scattering feature (CRSF) in the X-ray spectrum of M51 ULX-8 was discovered \citep{2018NatAs...2..312B,2019MNRAS.486....2M}. This feature which is directly related to the measurement of the magnetic field indicates the presence of a strong NS candidate in the ULX-8 system. Also very recently, the CRSF with the highest energy known to date was detected for the first Galactic PULX candidate Swift J0243.6+6124 by \cite{2022ApJ...933L...3K}.

Understanding the nature of the high luminosity mechanism and the compact objects (NSs or black holes, BHs) are the main purposes of ULXs studies. For this, X-ray-optical spectral and temporal analysis are very important tools. To determine possible optical counterpart(s) of ULXs, {\it Chandra} X-ray and {\it HST} optical observations which have high spatial resolution and positional accuracy usually are used. The number of detectable optical counterparts are quite limited since they are generally faint in the V-band (m$_{V}$ $\geq$ 21 mag). Moreover, they are located in the fields of crowded and star clusters. Therefore, the number of ULXs that have unique possible optical counterparts are not exceed 30 \citep{2011ApJ...737...81T,2013ApJS..206...14G}.

Optical analysis provides precious clues of the nature of the donor star and disk geometry. In addition, this is a very useful tool in determining the masses of compact objects \citep{2017ARA&A..55..303K}. However, the nature of optical emission of ULXs is still unclear. In other words, the optical emission could be originated from an accretion disk, a donor star or a combination of both. All these possibilities make it difficult for us to understand their optical nature. On the other hand, there have been many studies to identify and understand the nature of the possible optical counterpart(s) of ULXs. For instance, \cite{2002ApJ...580L..31L,2004MNRAS.351L..83K} and \cite{2005MNRAS.356...12S} reported that the high UV/optical magnitudes or blue colors of possible optical counterparts may be clues for OB-type stars. As a specific example, \cite{2014Natur.514..198M} showed that photospheric absorption lines were detected from the donor star in the blue part of the spectrum in NGC 7793 P13. However, \cite{2015NatPh..11..551F} pointed out that the He II line observed in NGC 7793 P13 must be formed in a photoionized wind from the accretion disc. On the other hand, the near-infrared counterparts of ULXs studies suggest that they may be red supergiants \citep{2016MNRAS.459..771H,2020MNRAS.497..917L}.
Moreover, the optical emission could be contaminated or dominated from an irradiated accretion disk \citep{2008MNRAS.386..543P,2012ApJ...745..123G,2014MNRAS.444.2415S,2019ApJ...884L...3Y}. 

The optical emission may originates from the photosphere of the wind at very high accretion rates with from irradiation of the donor star or together in some combination with the irradiated outer disk (\cite{2022MNRAS.509.1119M} and references therein). Moreover, for the first time the far-UV emission from the NGC 6946 ULX source was detected by \cite{2010ApJ...714L.167K} and they argued that it was a combination of disk emission and contribution from the donor star. In addition, Galactic super-Eddington accretor SS 433 which widely expected to be ultraluminous in the UV band \citep{2007MNRAS.377.1187P} has shown optical spectra which associated with the emission from the outer photosphere \citep{2007MNRAS.377.1187P}. In addition, the optical line luminosity limits combined with the X-ray variability may allow for beaming factors of a typically less than an order of magnitude. This is compatible with a model for supercritical accretion and large scale height winds \citep{2006MNRAS.370..399B}.\\

\begin{figure*}
\begin{center}
\includegraphics[angle=0,scale=0.40]{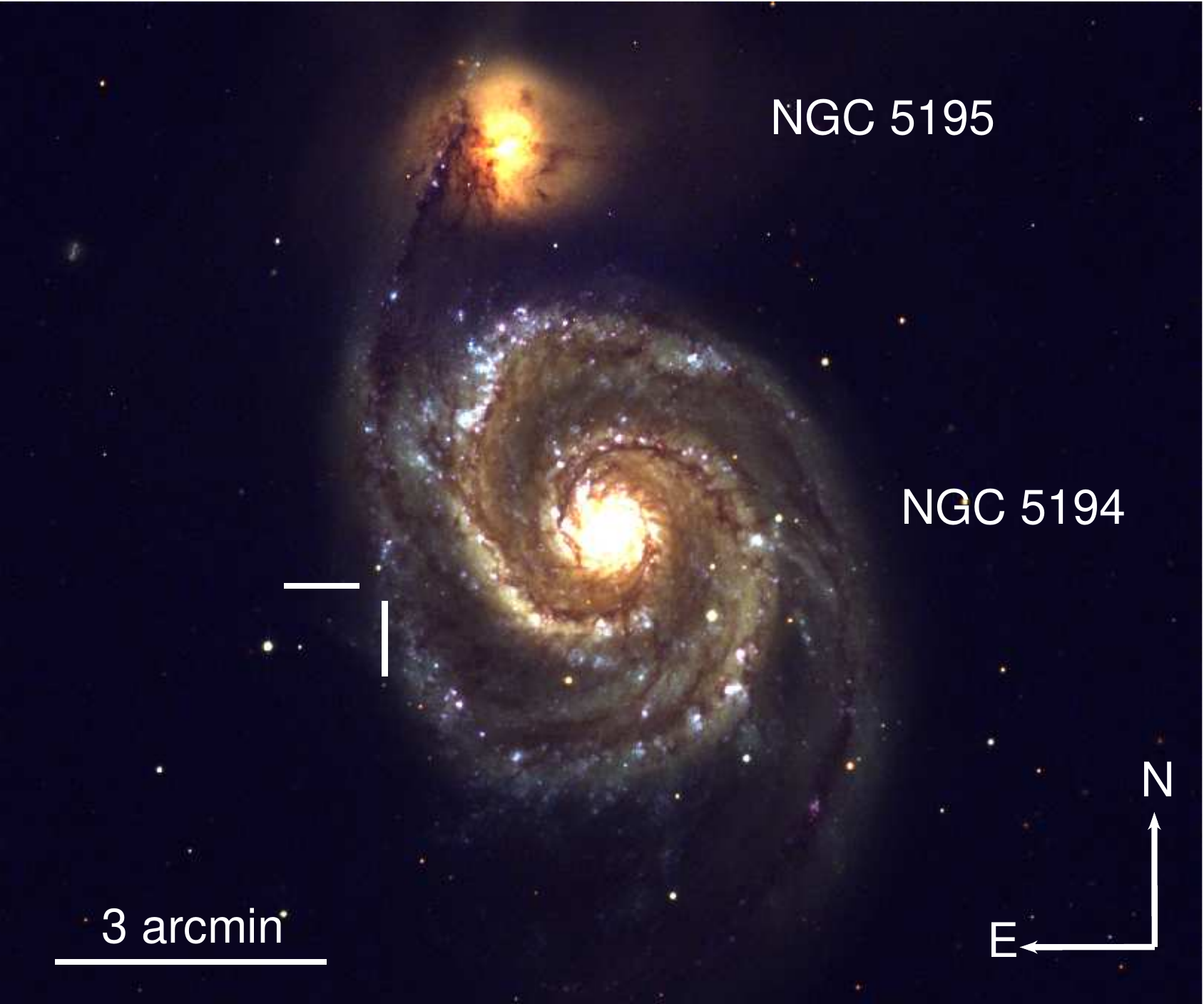}
\caption{RGB {\it SDSS} mosaic image of the M51 (NGC 5194/5194) galaxy. The white bars represent the position of ULX-8.}
\label{F:1}
\end{center}
\end{figure*}

The galaxy M51 (NGC 5194) and its companion NGC 5195 are a pair of interacting galaxies at a distance of 9 Mpc \citep{2020MNRAS.491.1260S}. This galaxy hosts a large number of ULXs \citep{2006ApJ...645..264T,2011ApJ...741...49S}. One of them is the well-known NS ULX-8. Figure \ref{F:1} shows the location of the ULX-8 on a three-color (red, green and blue; RGB) mosaic image of {\it SDSS} (Sloan Digital Sky Survey). This present work focuses on possible optical counterparts of M51 ULX-8. Primary goal is to search for potential optical counterparts for ULX-8 by deploying one-year {\it HST} ACS/WFC F606W and F814W observations. It is also an important part of this study to investigate in detail the nature of long-term optical and also X-ray variability.

The paper is organized as follows: in section \ref{sec:2}, data reduction and analysis (astrometric calculations, long-term optical variability, color magnitude diagrams and spectral energy distributions) are described. In section \ref{sec:3}, Results of the optical analysis, astrometry, long-term optical variability of counterparts and concluding main findings are presented.

\section{OBSERVATIONS, DATA REDUCTION AND ANALYSIS} \label{sec:2}
\subsection{Observations}

The M51 galaxy was observed many times by the {\it HST} and {\it Chandra} observatories. The main {\it HST} observations presented in this study are 34 drizzled images which were taken between 2016 and 2017 (Proposal ID 14704). These observations were obtained using the {\it HST} ACS F606W and F814W filters with the same exposure time of 2200 seconds (s). In addition, 2005-01-20 {\it HST}/ACS F435W (B-band), F555W (V-band) and F814W (I-band) drizzled images with exposure times of 2720 s, 1360 s, 1360 s, were also used to obtained for the color magnitude diagrams (CMDs) and spectral energy distributions (SEDs). Furthermore, archived 21 {\it Chandra} X-ray observations of ULX-8 (Table \ref{T:1}) were used for astrometric calculations as well as for X-ray temporal analysis.

\begin{table*}
\centering
\caption{{\it chandra} X-ray observations log for M51 ULX-8}
\begin{tabular}{ccccccccc}
\hline\hline
 ObsID & Instrument & Date & Exp. & F$_{X}$$^{a}$ & L$_{X}$$^{b}$ \\
 & & (YYYY-MM-DD) & (ks) &  (erg cm$^{-2}$ s$^{-1}$) & (erg s$^{-1}$)\\
\hline
354	    &	ACIS-S	&	2000-06-20	&	14.86	& 	3.79	$\pm$	0.21	&	3.67	$\pm$	0.20	\\
1622	&	ACIS-S	&	2001-06-23	&	26.81	&		2.28	$\pm$	0.12	&	2.21	$\pm$	0.12	\\
3932	&	ACIS-S	&	2003-08-07	&	47.97	&		3.25	$\pm$	0.12	&	3.15	$\pm$	0.12	\\
12562	&	ACIS-S	&	2011-06-12	&	9.63	&		6.42	$\pm$	0.39	&	6.22	$\pm$	0.38	\\
12668	&	ACIS-S	&	2011-07-03	&	9.99	&		3.34	$\pm$	0.27	&	3.24	$\pm$	0.26	\\
13813	&	ACIS-S	&	2012-09-09	&	179.20	& 	5.80	$\pm$	0.09	&	5.62	$\pm$	0.09	\\
13812	&	ACIS-S	&	2012-09-12	&	157.46	& 	4.19	$\pm$	0.08	&	4.06	$\pm$	0.08	\\
15496	&	ACIS-S	&	2012-09-19	&	40.97	& 	3.49	$\pm$	0.14	&	3.38	$\pm$	0.14	\\
13814	&	ACIS-S	&	2012-09-20	&	189.85	& 	3.20	$\pm$	0.06	&	3.10	$\pm$	0.06	\\
13815	&	ACIS-S	&	2012-09-23	&	67.18	& 	3.23	$\pm$	0.11	&	3.13	$\pm$	0.11	\\
13816	&	ACIS-S	&	2012-09-26	&	73.10	& 	3.80	$\pm$	0.11	&	3.68	$\pm$	0.11	\\
15553	&	ACIS-S	&	2012-10-10	&	37.57	& 	4.43	$\pm$	0.17	&	4.29	$\pm$	0.16	\\
{\it 19522}	&	ACIS-I	&	2017-03-17	&	37.76	& 		1.39	$\pm$	0.13	&	1.35	$\pm$	0.13	\\
{\it 20998}	&	ACIS-S	&	2018-08-31	&	19.82	& 	5.18	$\pm$	0.29	&	5.02	$\pm$	0.28	\\
{\it 23472}	&	ACIS-S	&	2020-10-13	&	33.62	& 	2.35	$\pm$	0.17	&	2.28	$\pm$	0.16	\\
{\it 23473}	&	ACIS-S	&	2020-11-18	&	34.51		&	2.67	$\pm$	0.19	&	2.59	$\pm$	0.18	\\
{\it 23474}	&	ACIS-S	&	2020-12-21	&	36.14		&	8.10	$\pm$	0.31	&	7.85	$\pm$	0.30	\\
{\it 23475}	&	ACIS-S	&	2021-01-28	&	34.51		&	8.86	$\pm$	0.35	&	8.58	$\pm$	0.34	\\
{\it 23476}	&	ACIS-S	&	2021-03-01	&	34.44		&	10.20	$\pm$	0.33	&	9.88	$\pm$	0.32	\\
{\it 23477}	&	ACIS-S	&	2021-04-01	&	31.64		&	3.38	$\pm$	0.21	&	3.27	$\pm$	0.20	\\
{\it 23478}	&	ACIS-S	&	2021-05-04	&	31.55		&	11.30	$\pm$	0.40	&	10.95	$\pm$	0.39	\\
\hline
\end{tabular}
\\Notes: $^{a}$ and $^{b}$ represent unabsorbed flux and luminosity in units of 10$^{-13}$ ergs $cm^{-2}$ $s^{-1}$ and 10$^{39}$ ergs $s^{-1}$ with 1-$\sigma$ errors, respectively. ObsIDs in italics indicate previously unused data for ULX-8.
\label{T:1}
\end{table*}

\subsection{Determination of Optical Counterparts}

In previous work, \cite{2006ApJ...645..264T} performed astrometry to determine the optical counterpart of M51 ULX8. In their work, point-like sources in the ACS images compared with sources in the 2MASS (Two Micron All Sky Survey) catalog. They found systematic offsets for R.A (right ascension) and for Decl. (declination) and they shifted the ACS images to the 2MASS positions. After this alignment, they determined four reference sources between ACS and {\it Chandra} images. As a result, they derived the astrometric error radius as 0$\arcsec$.3 at 90\% confidence between {\it Chandra} and {\it HST} images (see for more details \cite{2006ApJ...645..264T}.)

Using a different approach from the \cite{2006ApJ...645..264T}, the possible optical counterparts of the M51 ULX-8 were investigated as described follows: Optical counterparts of ULX sources could be determined by precise astrometric calculations using {\it HST} and {\it Chandra} observations, which generally have very good spatial resolution. Initially, the archive data sets from {\it HST} and {\it Chandra} were used to determine the possible optical counterparts of neutron star M51 ULX-8. For this, possible reference sources were searched by comparing 21 {\it Chandra} ACIS-S X-ray observations with {\it HST} ACS/F606W and F814W optical observations. The, {\it wavdetect} tool in {\scshape ciao} and the {\it daofind} task in {\it DAOPHOT}/{\scshape iraf} tasks were used for source detection in {\it Chandra} and {\it HST} observations, respectively. An align problem was noticed especially in {\it HST} ACS/F606W and F814W as well as {\it Chandra} observations. The {\it Chandra} observations were aligned by cross-matching another {\it Chandra} observation (ObsID 23478), which provided a relative correction. For these processes, {\it wcsmatch} and {\it wcsupdate} tools in {\scshape ciao} were used. The {\it imalign} tool in {\scshape iraf} was run to equalize the relatively large shifts in the W606W and F814W observations. However, for precision astrometry between {\it Chandra} and {\it HST} data sets, adequate reference sources were not found. Therefore, {\it GAIA} source catalog was used for astrometric calculations.

\begin{figure*}
\begin{center}
\includegraphics[angle=0,scale=0.55]{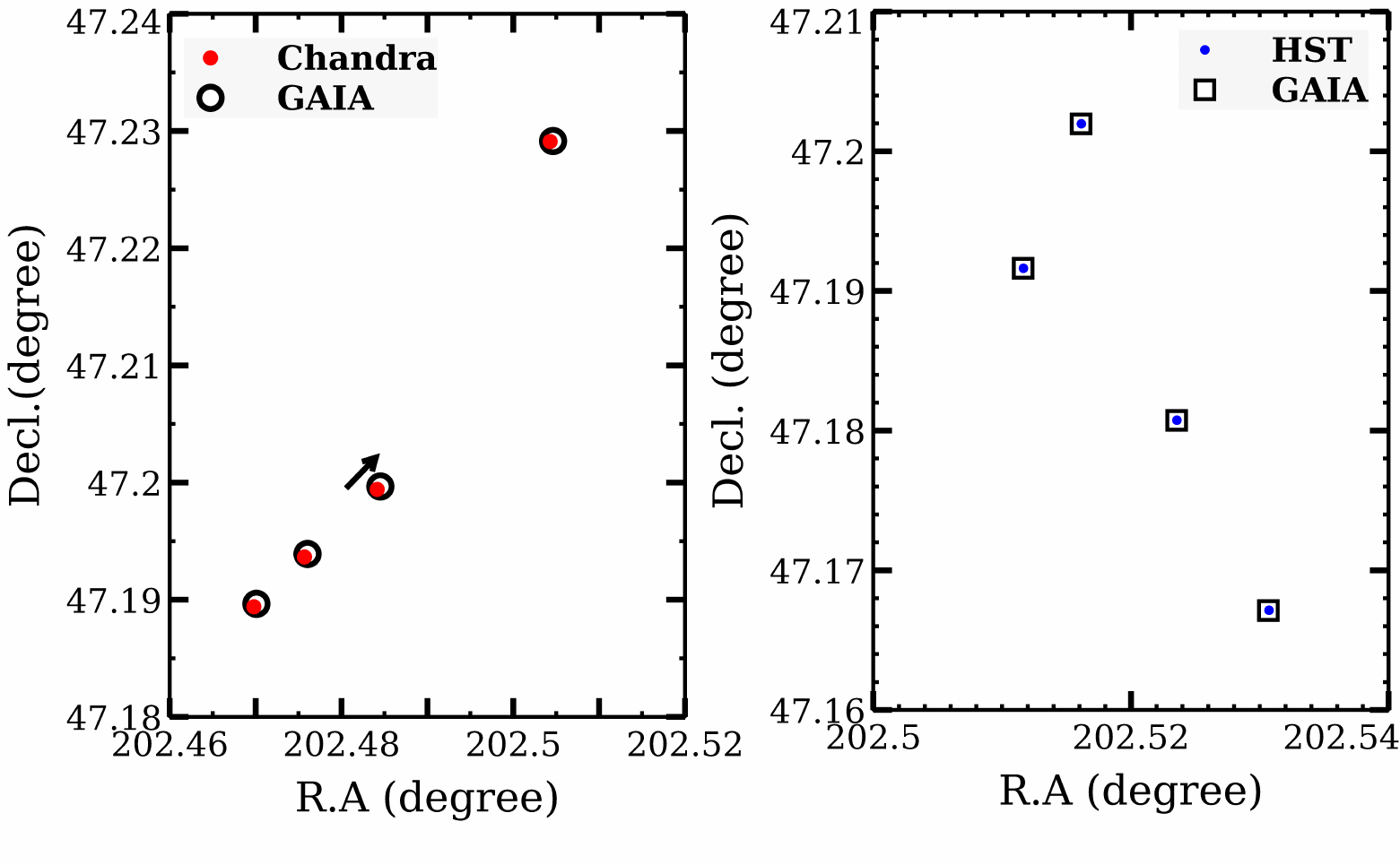}
\caption{Coordinates of {\it Chandra} \& {\it GAIA} (left) and {\it GAIA} \& {\it HST} (right) reference sources used in astrometric calculations. In the left panel, {\it Chandra} X-ray and {\it GAIA} optical reference sources are represented by filled red and unfilled black circles, respectively. The shift directions of the reference sources are indicated with the black arrow. In the right panel, {\it GAIA} and {\it HST} optical reference sources are represented by unfilled black squares and filled blue circles, respectively.}
\label{F:astro}
\end{center}
\end{figure*}

Since ds9 (An image display and visualization tool for astronomical data, \citealp{2003ASPC..295..489J}) already has an interface to various standard catalogues it was used to retrieve data from the {\it GAIA} catalogue. From this catalogue, the reference sources which appear to be isolated and point-like in {\it Chandra} and {\it HST} images were searched. Considering also the shift directions of matched sources, four reference sources were found between both {\it Chandra}-{\it GAIA} and {\it GAIA}-{\it HST}. Of course, there were more than four reference sources between {\it GAIA}-{\it HST}, but they were not affected the results since the values of shift as well as shift directions were the same. The coordinates of all identified reference sources are shown in Figure \ref{F:astro}. In particularly, the same shift direction of the reference sources between {\it Chandra} and {\it GAIA} indicates that the selected sources are the accurate references (see the left panel of Figure \ref{F:astro}). The astrometric offsets between {\it Chandra} and {\it GAIA} were found as -0$\arcsec$.83 $\pm$ 0$\arcsec$.07 for R.A and -0$\arcsec$.72 $\pm$ 0$\arcsec$.33 for Decl. with 1-$\sigma$ errors and also astrometric offsets between {\it GAIA} and {\it HST} were found as -0$\arcsec$.07 $\pm$ 0$\arcsec$.03 for R.A and -0$\arcsec$.04 $\pm$ 0$\arcsec$.02 for Decl. with 1-$\sigma$ errors. The total astrometric errors between {\it Chandra}-{\it GAIA} and between {\it GAIA} -{\it HST} were derived as 0$\arcsec$.33 and 0$\arcsec$.03, respectively. The calculation to determine the final optical coordinates of the ULX-8 is described in equations (1), (2) and (3):

\begin{equation}
  IC = IG \pm \Delta R.A_{(IC-IG)} \pm \Delta Decl._{(IC-IG)}
\end{equation}

\begin{equation}
  IG = IH \pm \Delta R.A_{(IG-IH)} \pm \Delta Decl._{(IG-IH)}
\end{equation}

Where the {\it Chandra}, {\it GAIA} and {\it HST} images are described as IC, IG and IH, respectively. $\Delta$R.A and $\Delta$Decl. are average shifts of R.A and Decl., respectively. The plus–minus signs, $\pm$, depend on the shift directions. To simplify, the last two terms of the equation (1) and equation (2) were called A and B respectively. The equation (3) gives the X-ray position of ULX-8 on the astrometric-corrected {\it HST} images.

\begin{equation}
     IC = IH \pm A \pm B
\end{equation}

\begin{figure*}
\begin{center}
\includegraphics[angle=0,scale=0.25]{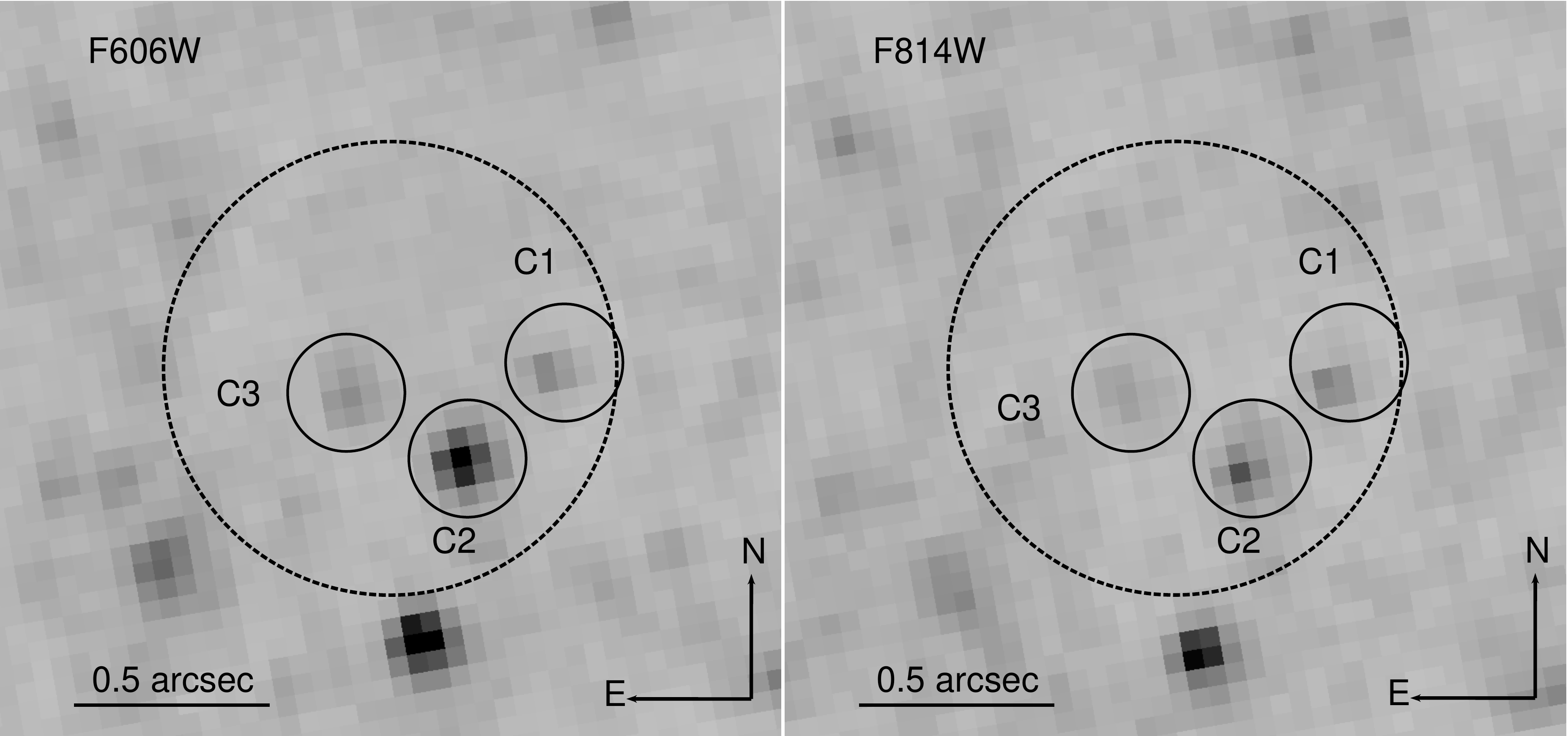}
\caption{The {\it Chandra} X-ray position of the M51 ULX-8 on the astrometric-corrected {\it HST/WFC} F606W (left) and F814W (right) images. Black dashed circles indicate the astrometric error radius of 0$\arcsec$.58 around the X-ray position and solid black circles represent the center coordinates of each possible optical counterpart (C1, C2 and C3).}
\label{F:can}
\end{center}
\end{figure*}

Finally, accepted 1-$\sigma$ accuracy in {\it Chandra} position as 0$\arcsec$.1, the positional error radius of ULX-8 was derived as 0$\arcsec$.58 at 90\% confidence level by combined all errors in quadrature. The {\it Chandra} X-ray position of the M51 ULX-8 on the astrometric-corrected {\it HST/WFC} F606W and F814W images are displayed in Figure \ref{F:can}. In the astrometric error radius, three discrete optical sources which have FWHM (The full Width at Half Maximum) as $\geq$ 5 pixels (0$\arcsec$.25) were identified. According to increasing R.A values of these sources are labeled as C1, C2 and C3 and the {\it HST} optical coordinates of possible optical counterparts are given in Table \ref{T:astro}.

\begin{table*}
\centering
\begin{minipage}[b]{0.9\linewidth}
\caption{The astrometric-corrected {\it Chandra} X-ray position of the M51 ULX-8 and Vega \\magnitudes of the possible optical counterparts C1, C2 and C3}
\begin{tabular}{ccccccccccr}
\hline
Source. &  {\it HST} R.A & {\it HST} Dec. &  F435W mag. & F555W mag. & F814W Mag.\\
 & (hh:mm:ss.sss) & ($\degr$ : $\arcmin$ : $\arcsec$) & (Vega)& (Vega) & (Vega)\\
\hline
ULX-8&  13:30:07.64 & +47:11:06.97 &...&...&...\\
C1 &   13:30:07.60 & +47:11:06.98  & 26.90 $\pm$ 0.18 &26.08 $\pm$ 0.17 & 25.72 $\pm$ 0.07\\
C2 &   13:30:07.62 & +47:11:06.76  & 23.90 $\pm$ 0.04 & 24.01 $\pm$ 0.06 & 24.77 $\pm$ 0.06\\
C3 &   13:30:07.65 & +47:11:06.94  & 26.17 $\pm$ 0.07 & 26.22 $\pm$ 0.06 & 26.03 $\pm$ 0.05\\
\hline
\end{tabular}
\label{T:astro}
\end{minipage}
\end{table*}

\subsection{Photometry of Possible Optical counterparts}

For all drizzled images, point-like sources were detected with the {\it daofind} task and aperture photometry of these sources was performed using the {\it APPHOT} package in {\scshape iraf}. To perform photometry, 3 pixels (0$\arcsec$.15) aperture radius was chosen. In order to obtain photometric errors and data quality propagation, the pixel values were multiplied by the exposure time using {\it imarith} tool in {\scshape iraf}. The Vega magnitudes were derived instrumental magnitudes using ACS/WFC zero point magnitudes from \cite{2005PASP..117.1049S}. To derive the aperture corrections with a radius between 3 pixels to 10 pixels for each image using 30 isolated and bright sources. The obtained Vega magnitude values were corrected with the foreground dust-extinction A$_{V}$ = 0.46 mag. from \cite{2022MNRAS.510.4355A}. The dereddened Vega magnitudes of the possible optical counterparts are given in Tables \ref{T:astro} and \ref{T:foto}.

In addition, the PSF (Point Spread Function) was performed to derive photometry following the similar approach given by \cite{2022MNRAS.510.4355A}. 30 bright and isolated sources near the optical counterparts were selected to build the PSF model using the {\it DAOPHOT} package
\cite{1987PASP...99..191S} in {\scshape iraf}. PSF fitting radius was taken as 3 pixels in the {\it allstar} task. The optical counterparts are discrete sources which are convenient for aperture photometry. In addition, in terms of their magnitudes there are no dramatic differences between the two methods therefore, only the results of aperture photometry are presented here.

\subsection{Color-Magnitude Diagrams \& Spectral Energy Distributions}

To classify possible optical counterparts, C1, C2 and C3, F$_{X}$/F$_{optik}$ ratios were derived using equation (5) which given by \cite{1982ApJ...253..504M}:

\begin{equation}
    log(F_X/F_{optik}) = log(F_X) + m_V/2.5 + 5.37
\end{equation}

Here, the F$_{X}$ is observed absorbed flux obtained from a simple {\it power-law} in the 0.3–3.5 keV energy range and m$_{V}$ is the V-band magnitude. Since there were no simultaneous {\it Chandra} and {\it HST} V-band observations, nearly simultaneous Mar. 17, 2017 (MJD: 57829.03334) {\it Chandra} and Mar. 21, 2017 (MJD: 57833.60502) {\it HST} F606W observations were used. F$_{X}$/F$_{optik}$ ratios of C1, C2 and C3 were derived as 2.58 $\pm$ 0.03, 1.65 $\pm$ 0.02 and 2.33 $\pm$ 0.03, respectively. The F$_{X}$/F$_{optik}$ value of C2 is in the range of AGNs (Active Galactic Nucleus) or BL Lac objects (-1 to 1.7) \citep{1991ApJS...76..813S}. 

\begin{figure}
\begin{center}
\includegraphics[width=\columnwidth]{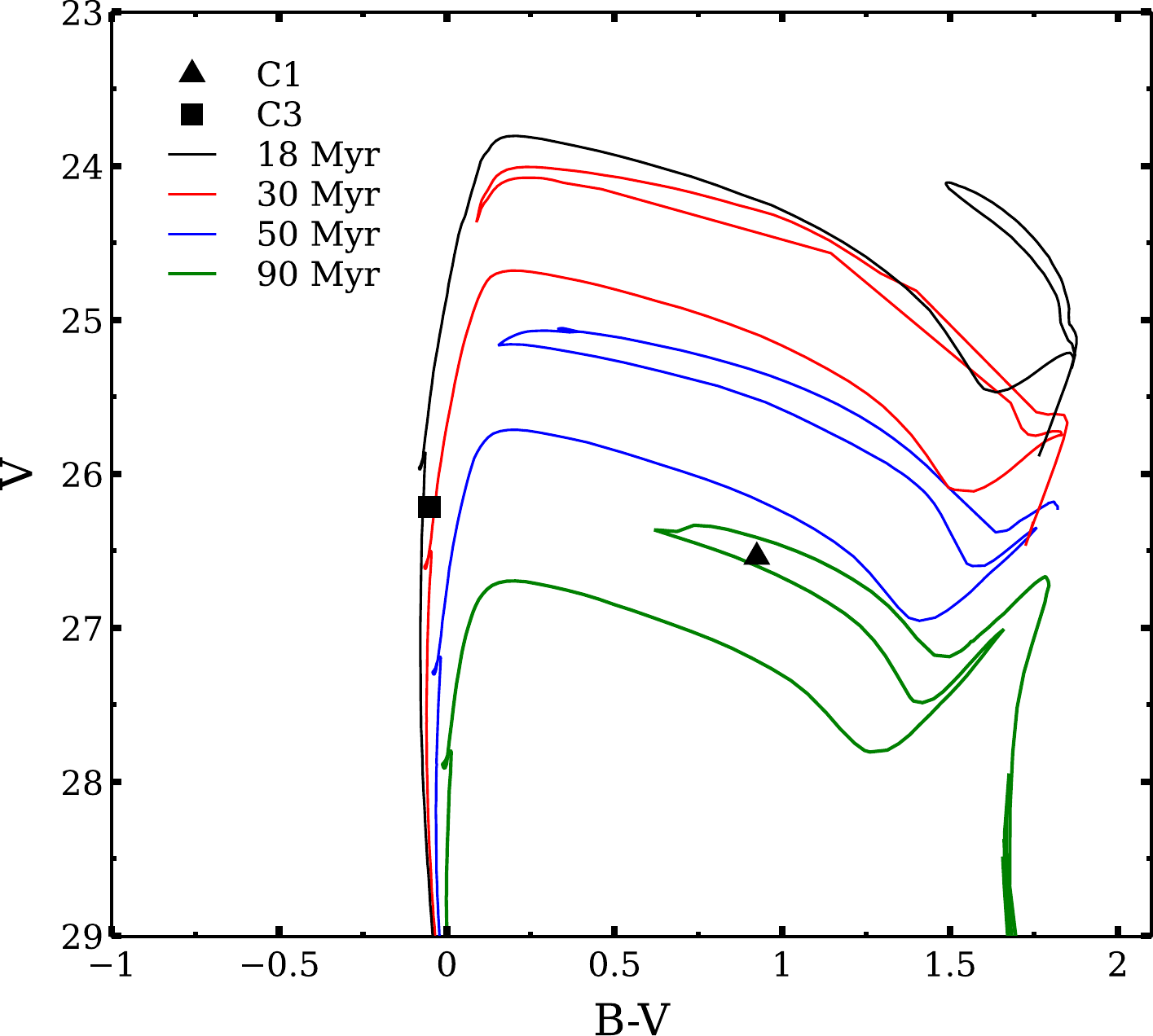}
\caption{CMD of ULX-8, the black, red, blue and green solid colours represent the age (18-90 Myr) isochrones of C1 (filled black triangle) and C3 (filled black square). The isochrones were corrected for extinction of A$_{V}$ $=$ 0.46 mag.}
\label{F:CMD}
\end{center}
\end{figure}

Investigating the environment (e.g. nearby stars or group of stars) of the ULX-8 could be a good tool to estimate the ages of its possible optical counterparts. In order to constrain the ages of the possible optical counterparts, CMDs were obtained using the colours of nearby stars of ULX-8 and counterparts. For this, similar colour values of possible optical counterparts and field stars were chosen. CMD, V-band versus B$-$V colours were derived for optical counterparts and the nearby field stars. PARSEC \citep{2012MNRAS.427..127B} isochrones were generated using the metallicity of Z=0.015 \citep{2018MNRAS.475.3561U} and A$_{V}$=0.46 mag. The distance modulus was derived as 29.7 mag. using a distance of 9 Mpc. The CMD of C1 and C3 are displayed in Figure \ref{F:CMD}. According to resultant CMD, the age ranges of C1 (black triangle) and C3 (black square) could be constrained to $\sim$ 90 Myr for C1 and (18-30) Myr for C3.

\begin{figure}
\begin{center}
\includegraphics[width=\columnwidth]{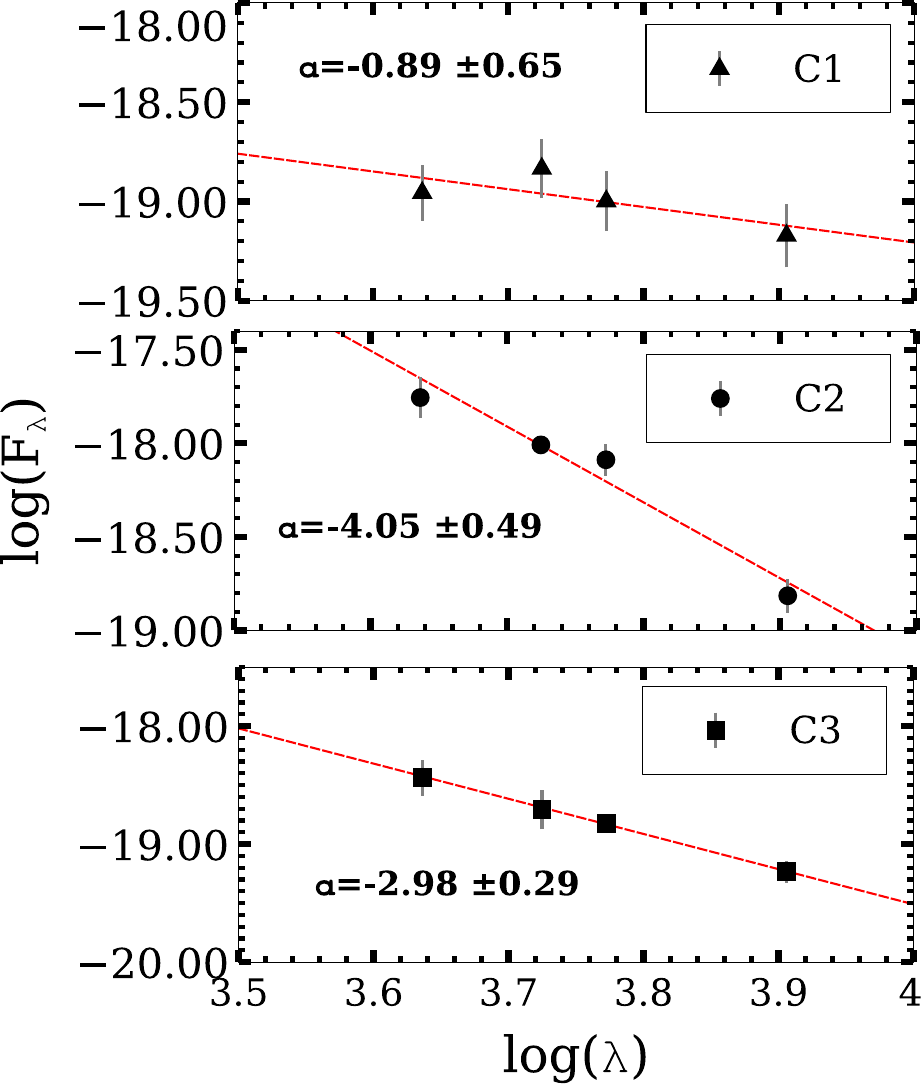}\caption{The reddening corrected SED of the possible counterparts of ULX-8. The {\it power-law} model is shown by dashed red lines. The source labels and model parameter are also denoted in the captions of each plot.The units of y and x axes are erg cm$^{-2}$ s$^{-1}$ \AA$^{-1}$ and \AA, respectively.}
\label{F:SEDs}
\end{center}
\end{figure}

Spectral energy distributions (SEDs) of the possible optical counterparts constructed to obtain the possible spectral characteristics of ULX-8 using the derived flux values B-band, V-band, I-band and also F606W filter. For SED plots, the pivot wavelengths of the ACS/WFC filters were selected. The optical SED for only counterpart C3 was adequately well-fitted with {\it power-law} model with a photon index, $\alpha$$=$$-$ 2.98 $\pm$ 0.29 at 95\% confidence level. However, an acceptable model could not be obtained for the remaining C2 and C3. The $\chi^2_{\nu}$ of {\it power-law} model for C1, C2 and C3 were found as 0.49, 1.89 and 0.84, respectively. In all cases, the number of degrees of freedom (dof) is two. The SEDs of all possible counterparts are displayed in Figure \ref{F:SEDs}. 

\subsection{Temporal Analysis of Possible Optical Counterparts}

\begin{figure*}
\begin{center}
\includegraphics[angle=0,scale=0.33]{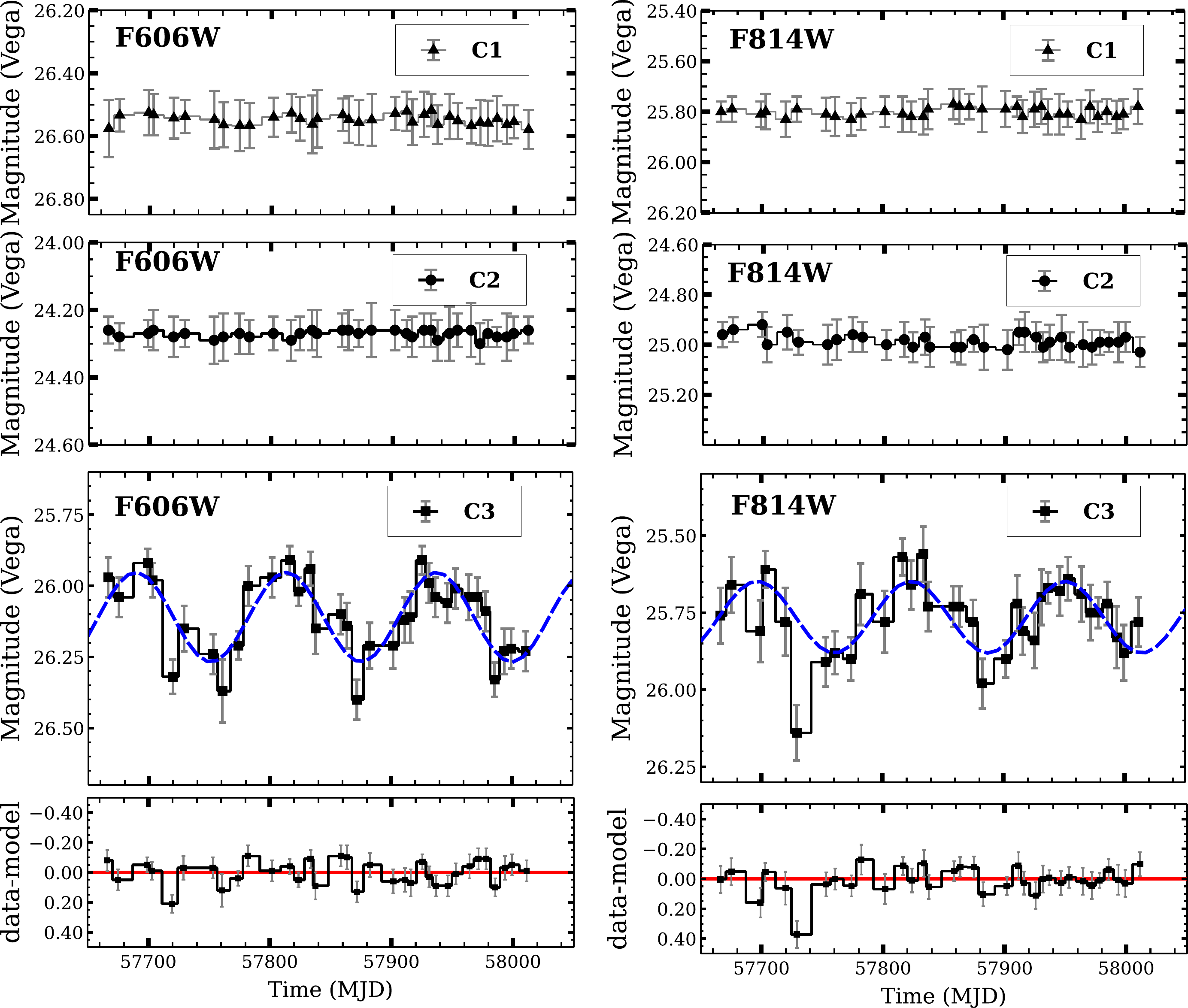}
\caption{The light-curves of the possible optical counterparts C1 (upper), C2 (center) and C3 (lower) in the F606W and the F814W filters. For counterpart C3, sinusoidal fits are represented blue dashed lines and their residuals are shown the lower panels.}
\label{F:period}
\end{center}
\end{figure*}

The Vega magnitudes and associated and 1-$\sigma$ errors of possible optical counterparts of the ULX-8 in the F606W and F814W filters were listed in Table \ref{T:foto}.
In order to search for periodic variability of possible optical counterparts, the long-term light-curves were produced for each filter using all ACS/WFC observations. As seen in the light-curves in the lower panel of Figure \ref{F:period}, a sinusoidal shape was seen for candidate C3 albeit it was not observed for the C1 and C2. The standard deviation values ($\leq$0.09) of the C1 and C2 magnitudes are almost equivalent to their errors, this indicates that these sources do not have optical variations in the F606W and F814W filters. The F606W and F814W light-curves of C3 well-fitted a sinusoidal curve of equation (4) given by \cite{2009ApJ...690L..39L}:

\begin{equation} \label{E:4}
     m(t) = \overline{m} + {\it A}sin[2\pi({\it t}-{\it t_{1}})/{\it P} + \phi]
\end{equation}\\

From the best fitting of the F606W light-curve, a period ({\it P}) was found as 124.12 $\pm$ 0.27 days (d) with amplitude {\it A}= 0.14 $\pm$ 0.02 mag. at 95\% confidence level. In case of F814W observations, the {\it P} and {\it A} were found as 127.04 $\pm$ 2.24 d and 0.11 $\pm$ 0.02, respectively, at 95\% confidence level. The reduced $\chi^{2}$ values of the fitting F606W and F814W light-curves were 1.48 and 1.27, respectively. In both cases, the dof was 31. The well-fitted parameters ({\it P}, {\it A}, $\overline{m}$ (average magnitude) and $\phi$ (phase)) of both light-curves are given in Table \ref{T:fit}.

\begin{table*}
\centering
\begin{minipage}[b]{0.9\linewidth}
\caption{Vega magnitudes of the possible optical counterparts C1, C2 and C3}
\begin{tabular}{ccccccccccr}
\hline
& \multicolumn{3}{c} {F606W Vega Magnitudes} && \multicolumn{3}{c} {F814W Vega Magnitudes}\\

Number & Time & C1 & C2 & C3 & Time & C1 & C2 & C3 \\
(1)& (2) & (3) & (4) & (5) & (6) & (7) & (8) & (9)\\
\hline
1	&	57666.23008	&	26.58	$\pm$ 0.07	&	24.26	$\pm$	0.04	&	25.97	$\pm$	0.07	&	57666.29509	&	25.80	$\pm$	0.04	&	24.96	$\pm$	0.05	&	25.76	$\pm$	0.09	\\
2	&	57675.23803	&	26.53	$\pm$	0.09	&	24.28	$\pm$	0.04	&	26.04	$\pm$	0.07	&	57675.30338	&	25.79	$\pm$	0.05	&	24.94	$\pm$	0.05	&	25.66	$\pm$	0.09	\\
3	&	57699.07316	&	26.53	$\pm$	0.05	&	24.27	$\pm$	0.04	&	25.92	$\pm$	0.05	&	57699.13777	&	25.81	$\pm$	0.05	&	24.92	$\pm$	0.05	&	25.81	$\pm$	0.10	\\
4	&	57703.11242	&	26.53	$\pm$	0.07	&	24.26	$\pm$	0.06	&	25.98	$\pm$	0.06	&	57703.17599	&	25.80	$\pm$	0.07	&	25.00	$\pm$	0.07	&	25.61	$\pm$	0.06	\\
5	&	57719.68466	&	26.54	$\pm$	0.07	&	24.28	$\pm$	0.06	&	26.32	$\pm$	0.06	&	57719.77352	&	25.83	$\pm$	0.07	&	24.95	$\pm$	0.07	&	25.78	$\pm$	0.11	\\
6	&	57729.02204	&	26.54	$\pm$	0.07	&	24.27	$\pm$	0.04	&	26.15	$\pm$	0.08	&	57729.10245	&	25.79	$\pm$	0.05	&	24.99	$\pm$	0.05	&	26.14	$\pm$	0.09	\\
7	&	57753.05545	&	26.55	$\pm$	0.05	&	24.29	$\pm$	0.07	&	26.24	$\pm$	0.07	&	57753.12005	&	25.81	$\pm$	0.07	&	25.00	$\pm$	0.08	&	25.91	$\pm$	0.08	\\
8	&	57760.60783	&	26.56	$\pm$	0.09	&	24.28	$\pm$	0.07	&	26.37	$\pm$	0.11	&	57760.68225	&	25.82	$\pm$	0.08	&	24.98	$\pm$	0.08	&	25.88	$\pm$	0.07	\\
9	&	57773.93019	&	26.57	$\pm$	0.07	&	24.27	$\pm$	0.06	&	26.21	$\pm$	0.05	&	57773.99199	&	25.83	$\pm$	0.06	&	24.96	$\pm$	0.07	&	25.90	$\pm$	0.07	\\
10	&	57782.01584	&	26.57	$\pm$	0.08	&	24.28	$\pm$	0.05	&	26.00	$\pm$	0.07	&	57782.09514	&	25.81	$\pm$	0.06	&	24.97	$\pm$	0.06	&	25.69	$\pm$	0.10	\\
11	&	57801.73972	&	26.54	$\pm$	0.07	&	24.27	$\pm$	0.05	&	25.97	$\pm$	0.07	&	57801.80433	&	25.80	$\pm$	0.06	&	25.00	$\pm$	0.06	&	25.78	$\pm$	0.10	\\
12	&	57816.38990	&	26.53	$\pm$	0.06	&	24.29	$\pm$	0.06	&	25.91	$\pm$	0.05	&	57816.50321	&	25.81	$\pm$	0.07	&	24.98	$\pm$	0.07	&	25.57	$\pm$	0.06	\\
13	&	57823.60626	&	26.55	$\pm$	0.06	&	24.27	$\pm$	0.05	&	26.02	$\pm$	0.05	&	57823.67499	&	25.82	$\pm$	0.06	&	25.01	$\pm$	0.06	&	25.66	$\pm$	0.08	\\
14	&	57833.60502	&	26.56	$\pm$	0.07	&	24.26	$\pm$	0.06	&	25.94	$\pm$	0.06	&	57833.66838	&	25.82	$\pm$	0.07	&	24.97	$\pm$	0.07	&	25.56	$\pm$	0.09	\\
15	&	57837.64349	&	26.54	$\pm$	0.09	&	24.27	$\pm$	0.07	&	26.15	$\pm$	0.09	&	57837.70811	&	25.79	$\pm$	0.08	&	25.01	$\pm$	0.08	&	25.73	$\pm$	0.08	\\
16	&	57858.50069	&	26.53	$\pm$	0.09	&	24.26	$\pm$	0.05	&	26.10	$\pm$	0.07	&	57858.56530	&	25.77	$\pm$	0.06	&	25.01	$\pm$	0.06	&	25.73	$\pm$	0.07	\\
17	&	57863.53201	&	26.55	$\pm$	0.05	&	24.26	$\pm$	0.06	&	26.14	$\pm$	0.08	&	57863.59661	&	25.78	$\pm$	0.07	&	25.01	$\pm$	0.07	&	25.73	$\pm$	0.07	\\
18	&	57871.80715	&	26.56	$\pm$	0.07	&	24.27	$\pm$	0.04	&	26.40	$\pm$	0.07	&	57871.87212	&	25.78	$\pm$	0.05	&	24.98	$\pm$	0.05	&	25.78	$\pm$	0.07	\\
19	&	57882.35745	&	26.55	$\pm$	0.07	&	24.26	$\pm$	0.08	&	26.21	$\pm$	0.08	&	57882.43069	&	25.79	$\pm$	0.09	&	25.01	$\pm$	0.09	&	25.98	$\pm$	0.08	\\
20	&	57901.68632	&	26.53	$\pm$	0.08	&	24.26	$\pm$	0.06	&	26.21	$\pm$	0.08	&	57901.75265	&	25.79	$\pm$	0.07	&	25.02	$\pm$	0.08	&	25.90	$\pm$	0.06	\\
21	&	57911.21975	&	26.52	$\pm$	0.05	&	24.27	$\pm$	0.04	&	26.12	$\pm$	0.08	&	57911.30027	&	25.78	$\pm$	0.04	&	24.95	$\pm$	0.05	&	25.72	$\pm$	0.09	\\
22	&	57915.85382	&	26.56	$\pm$	0.06	&	24.28	$\pm$	0.06	&	26.11	$\pm$	0.09	&	57915.91841	&	25.82	$\pm$	0.07	&	24.95	$\pm$	0.08	&	25.81	$\pm$	0.08	\\
23	&	57925.58697	&	26.53	$\pm$	0.07	&	24.26	$\pm$	0.05	&	25.91	$\pm$	0.05	&	57925.64973	&	25.79	$\pm$	0.07	&	24.97	$\pm$	0.06	&	25.84	$\pm$	0.09	\\
24	&	57931.34880	&	26.52	$\pm$	0.07	&	24.26	$\pm$	0.05	&	25.99	$\pm$	0.07	&	57931.41506	&	25.78	$\pm$	0.07	&	25.01	$\pm$	0.06	&	25.70	$\pm$	0.09	\\
25	&	57936.61826	&	26.56	$\pm$	0.05	&	24.29	$\pm$	0.06	&	26.04	$\pm$	0.07	&	57936.73306	&	25.82	$\pm$	0.07	&	24.99	$\pm$	0.07	&	25.67	$\pm$	0.05	\\
26	&	57946.45948	&	26.54	$\pm$	0.06	&	24.27	$\pm$	0.08	&	26.06	$\pm$	0.07	&	57946.52410	&	25.81	$\pm$	0.08	&	24.97	$\pm$	0.09	&	25.68	$\pm$	0.08	\\
27	&	57953.14623	&	26.55	$\pm$	0.07	&	24.26	$\pm$	0.05	&	26.01	$\pm$	0.07	&	57953.21083	&	25.81	$\pm$	0.05	&	25.01	$\pm$	0.06	&	25.64	$\pm$	0.07	\\
28	&	57964.26810	&	26.57	$\pm$	0.06	&	24.26	$\pm$	0.08	&	26.04	$\pm$	0.08	&	57964.33258	&	25.83	$\pm$	0.08	&	25.00	$\pm$	0.09	&	25.69	$\pm$	0.09	\\
29	&	57971.61629	&	26.56	$\pm$	0.06	&	24.30	$\pm$	0.06	&	26.04	$\pm$	0.07	&	57971.68077	&	25.78	$\pm$	0.07	&	25.01	$\pm$	0.07	&	25.75	$\pm$	0.09	\\
30	&	57978.10350	&	26.56	$\pm$	0.07	&	24.27	$\pm$	0.04	&	26.09	$\pm$	0.07	&	57978.16831	&	25.82	$\pm$	0.06	&	24.99	$\pm$	0.05	&	25.75	$\pm$	0.05	\\
31	&	57985.54818	&	26.54	$\pm$	0.07	&	24.28	$\pm$	0.03	&	26.33	$\pm$	0.06	&	57985.61031	&	25.80	$\pm$	0.05	&	24.99	$\pm$	0.04	&	25.72	$\pm$	0.07	\\
32	&	57993.48529	&	26.56	$\pm$	0.07	&	24.28	$\pm$	0.07	&	26.23	$\pm$	0.08	&	57993.55398	&	25.82	$\pm$	0.06	&	24.99	$\pm$	0.08	&	25.83	$\pm$	0.10	\\
33	&	57999.30881	&	26.55	$\pm$	0.06	&	24.27	$\pm$	0.05	&	26.22	$\pm$	0.07	&	57999.37343	&	25.81	$\pm$	0.06	&	24.97	$\pm$	0.06	&	25.88	$\pm$	0.09	\\
34	&	58011.35797	&	26.58	$\pm$	0.05	&	24.26	$\pm$	0.04	&	26.23	$\pm$	0.07	&	58011.42352	&	25.78	$\pm$	0.07	&	25.03	$\pm$	0.06	&	25.78	$\pm$	0.08	\\
\hline
\end{tabular}
\\ Notes: (1) shows the number of observations. (2) and (6) are MJD (Modified Julian Day) times of F606W and F814W observations, respectively. (3), (4), (5), (7), (8) and (9) show F606W and F814W Vega magnitudes of possible optical counterparts, respectively. \\
\label{T:foto}
\end{minipage}
\end{table*}

\begin{table*}
\centering
\begin{minipage}[b]{0.9\linewidth}
\caption{Best sinusoidal fitting parameters of possible optical counterpart C3.}
\begin{tabular}{cccccccc}
\hline\hline
Parameters & Units & F606W & F814W \\
\hline
{\it P} & Period (days) & 124.12 $\pm$ 0.27 & 127.04 $\pm$ 2.24 \\
{\it A} & Amplitude (mag.)& 0.16 $\pm$ 0.01 & 0.11 $\pm$ 0.02 \\
{\it $\phi$} & Phase ($^{\circ}$) & 217.18 $\pm$ 0.11 & 216.85 $\pm$ 0.15 \\
$\overline{m}$ & Average magnitude & 26.11  & 25.77  \\
$\chi^{2}$ & Chi-square test & 45.99 & 39.51 \\
{\it dof} & Number of degrees of freedom & 31 & 31 \\
$\chi^2_{\nu}$ & $\chi^{2}$/{\it dof} & 1.48 & 1.27 \\
\hline
\end{tabular}
\label{T:fit}
\end{minipage}
\end{table*}

\begin{figure*}
\begin{center}
\includegraphics[angle=0,scale=0.33]{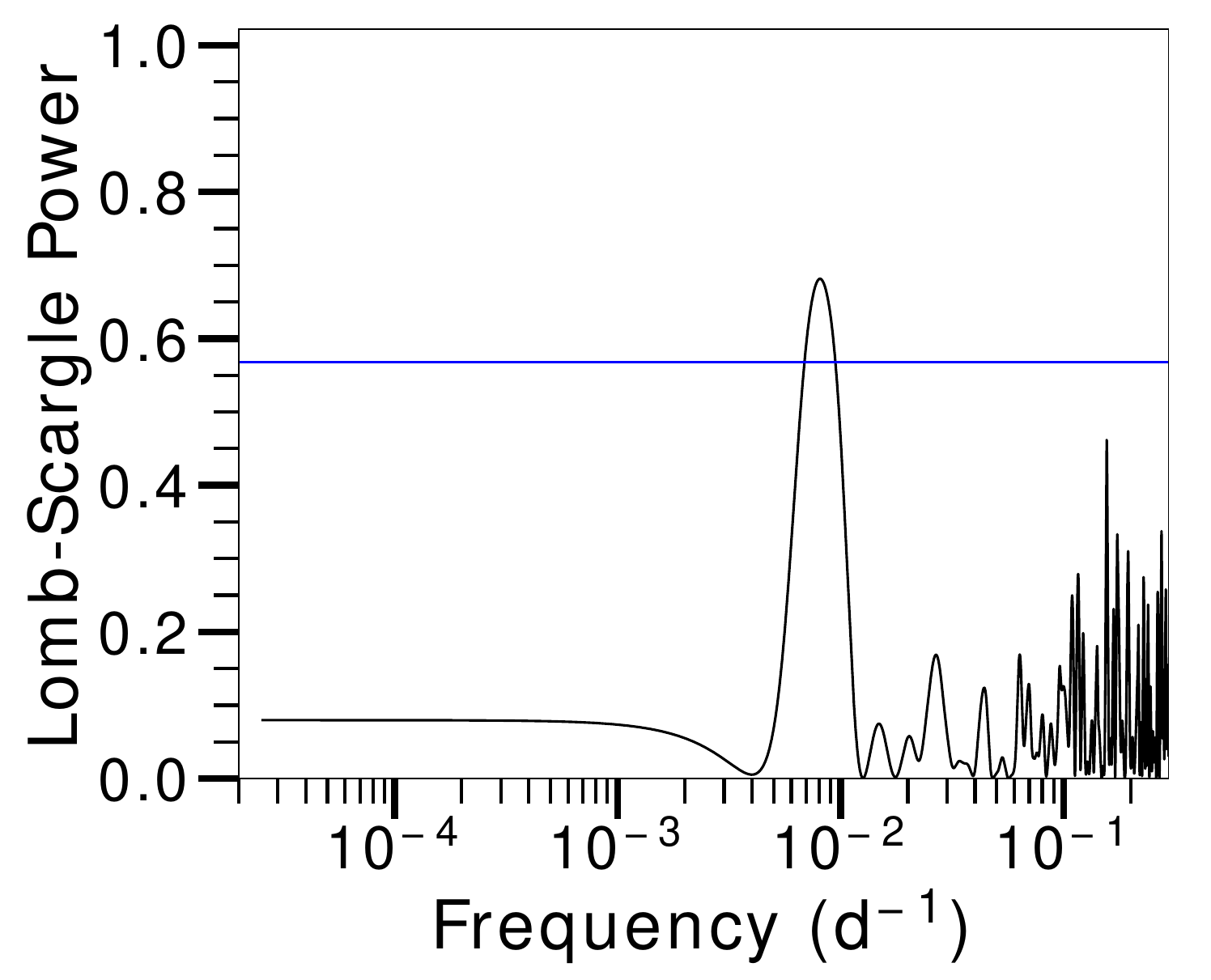}
\includegraphics[angle=0,scale=0.33]{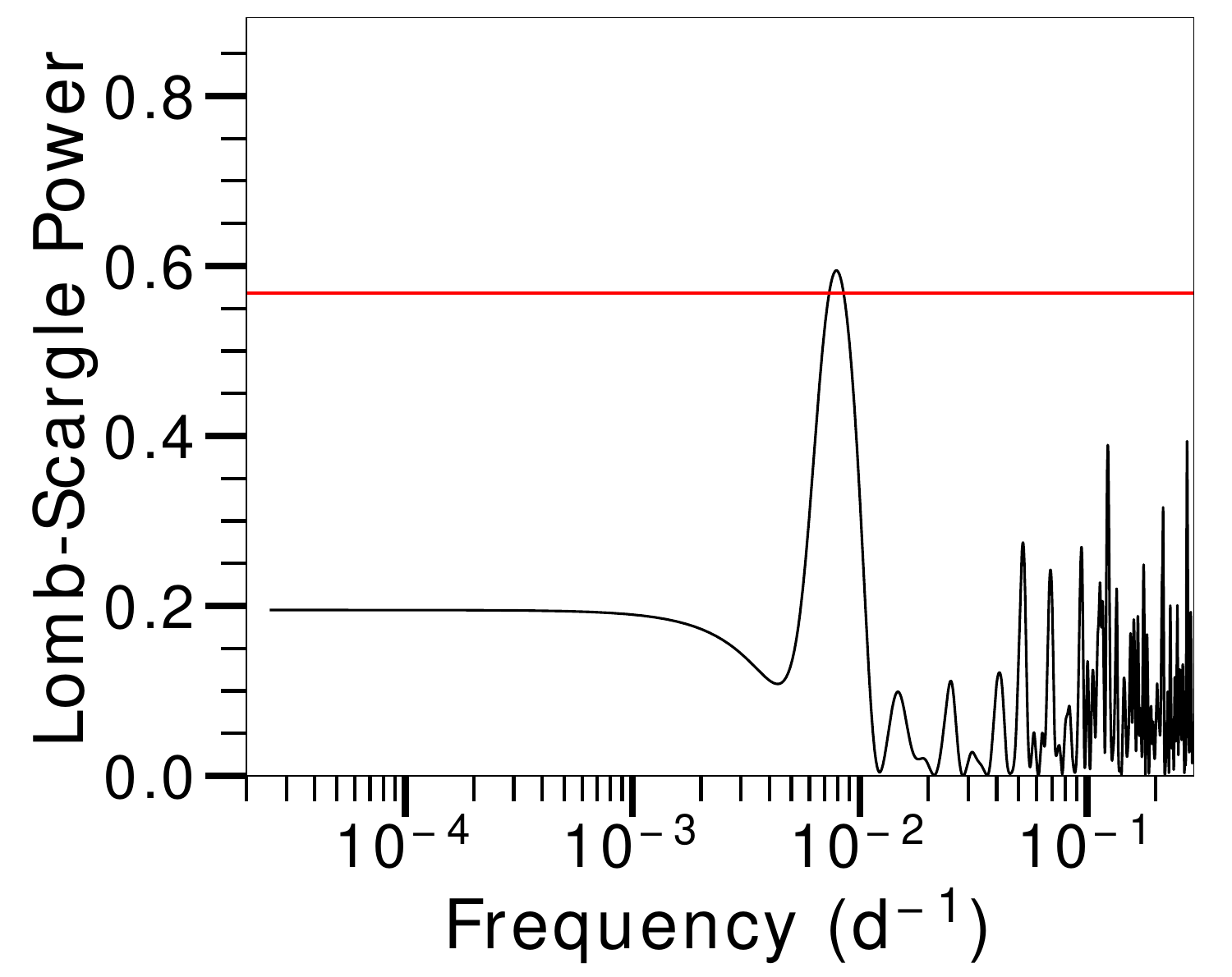}
\caption{Lomb-Scargle periodograms of the F606W (left) and F814W (right) observations. Periodic signals that peak at 124 d and 127 d are found for the F606W (left)and F814W, respectively. The solid blue and red lines represent the 95\% confidence level.}
\label{F:ls}
\end{center}
\end{figure*}

Moreover, to check long-term periodicity, Lomb-Scargle (L-S) periodogram \citep{1976Ap&SS..39..447L,1982ApJ...263..835S} which is a sub-package time series of the {\it Python/Astropy} was used. From the L-S periodogram, $\sim$ 124 d and $\sim$ 127 d periods were found with false alarm probability (FAP) as 3.1$\times$10$^{-3}$ and 2.8$\times$10$^{-2}$, respectively (see Figure \ref{F:ls}). In order to estimate the uncertainty on the periods a Monte Carlo simulation was used with {\it astropy uncertainty package}. Following the similar approach given by \cite{2010MNRAS.409.1220D} the 0.58 and 3.66 d at 95\% confidence level errors were found for the 124 and 127 d periods of ULX-8, respectively. The final period and amplitude of C3 were derived as 125.5 $\pm$ 2.3 d and 0.14 $\pm$ 0.11 by averaged the values which were found from the sinusoidal curve and the L-S analysis.

Numerous X-ray binaries and AGNs may show persistent random modulations such as stochastic and red-noise since in their emission driven by the complex and turbulent accretion process \citep{2016MNRAS.461.3145V}. In order to search the probabilities of a false detection due to red-noise, the Monte-Carlo simulation was used following the procedure used by \cite{2013MNRAS.434.3487A,2016A&A...585A..89A}. A total of 10000 simulated red-noise time series which have the same number of sampling points and the same parameters given in Table \ref{T:fit} were generated for each filter. This allows us to estimate high significance against possible false detection. To estimate the red-noise continuum the L-S is applied for each simulated light-curve with the {\it power-law} index ranging between -1 and -2 and with the multi-frequency. The simulated maximum power was not exceed the maximum power from the real data at 99.7\% confidence level. This result supports the interpretation that the determined periodicity is not spurious. Moreover, the light-curves of all the optical counterparts are different this may indicates that the periodic modulation is not due to systematic error and the effects of background images.\\

\begin{figure}
\begin{center}
\includegraphics[width=\columnwidth]{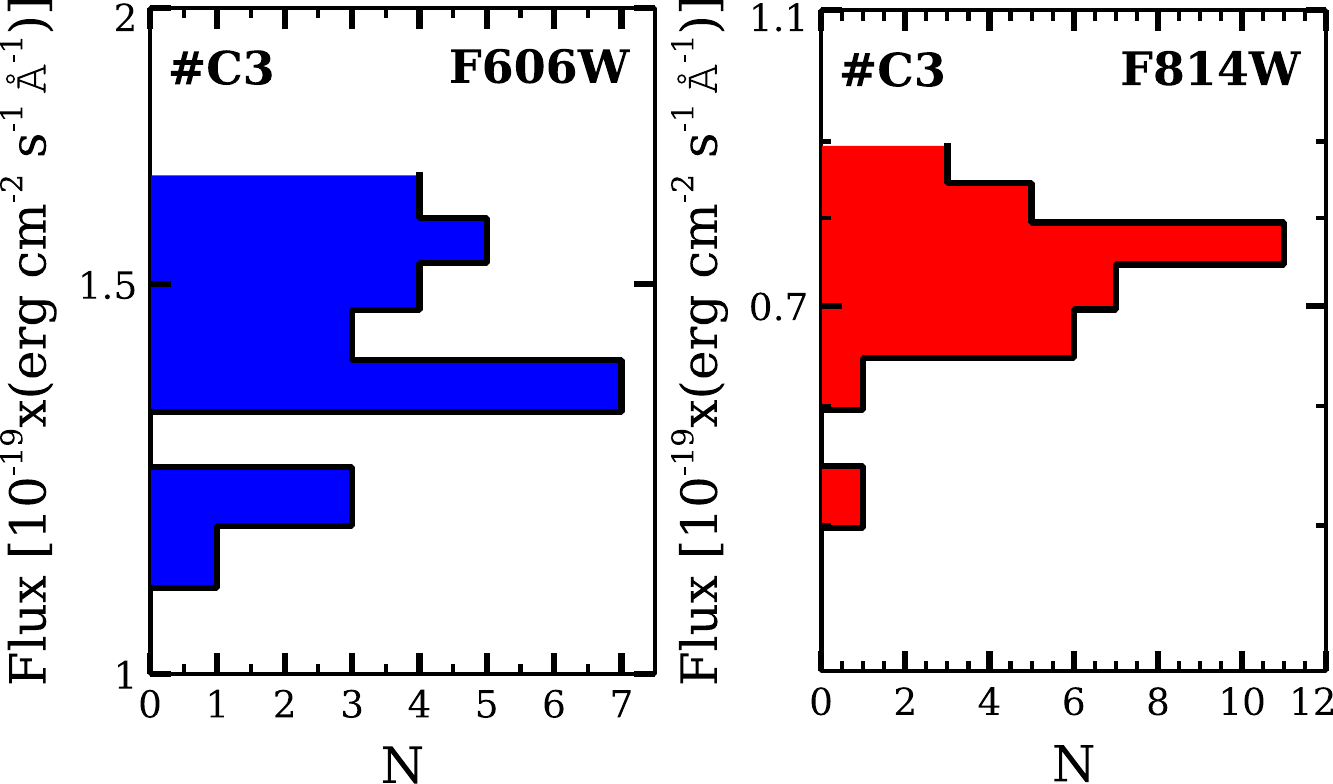}
\caption{The histogram of fluxes F606W (left) and F814W (right) observations for possible counterpart C3. Observations binned with 12 days and N is the number of observations.}
\label{F:histo}
\end{center}
\end{figure}

In addition, the observed optical flux distributions for source C3 are displayed in Figure \ref{F:histo} using F606W (left) and F814W (right) observations. As seen in this figure, the source C3 showed evidence of bi-modal distributions. The fluxes are derived from the Vega magnitudes of both filters. For the flux calculations, pivot wavelengths and the flux densities were taken from the \cite{2005PASP..117.1049S}.

\begin{figure}
\begin{center}
\includegraphics[width=\columnwidth]{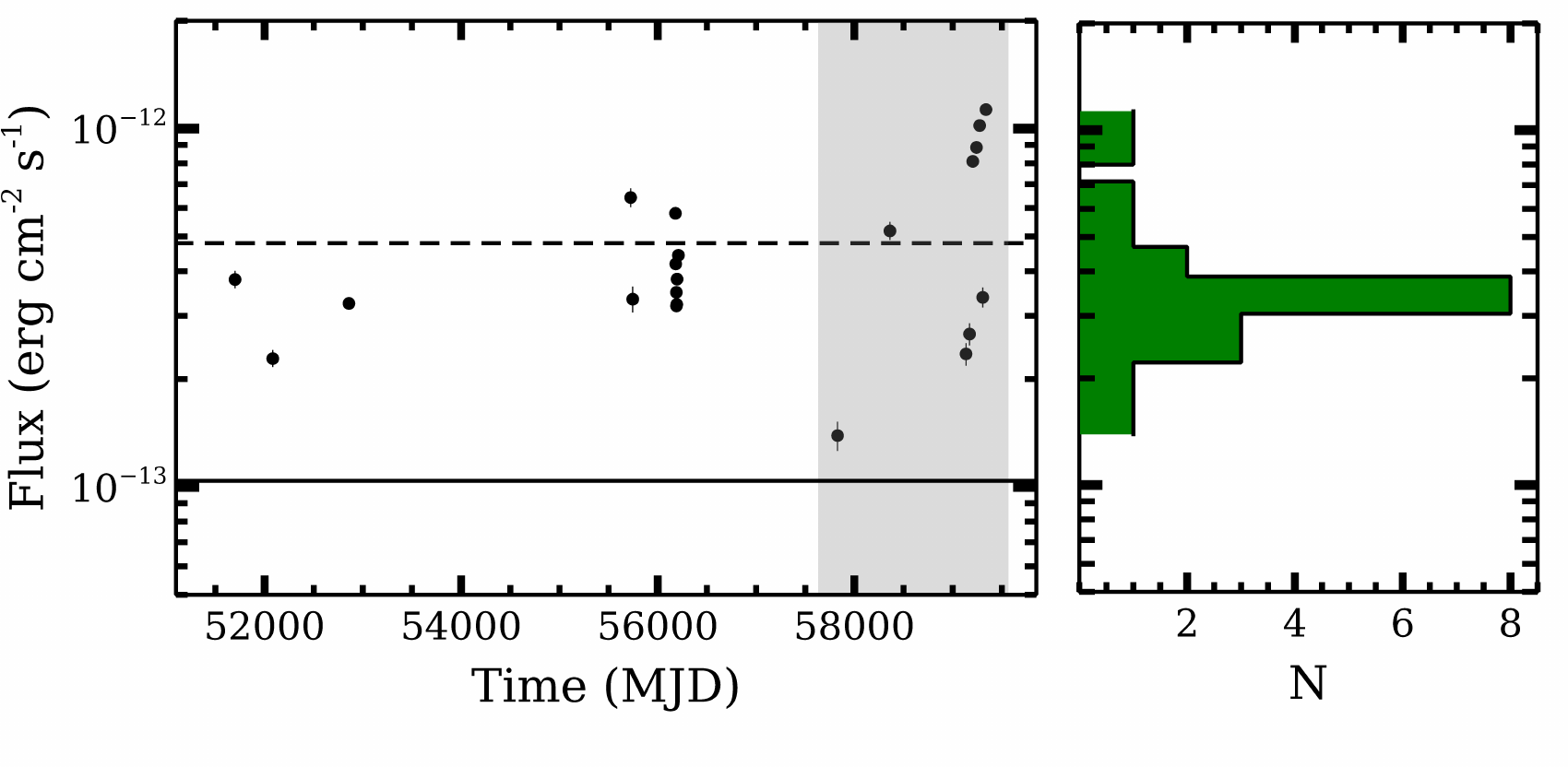}
\caption{The long-term {\it Chandra} X-ray light-curve of M51 ULX-8 (left) and its histogram (right). The histogram of fluxes was binned with 12 d. N is the number of observations. Dashed and solid black lines indicate the average of the fluxes and the flux threshold for ULX state adopted distance of 9 Mpc, respectively. The gray colored area represents previously unused observations for the ULX-8.}
\label{F:xraylc}
\end{center}
\end{figure}

Recently, new {\it Chandra} X-ray observations of galaxy M51 have been published (e.g. \cite{2020cxo..prop.5840E}). To quick look at the nature of X-ray variability of ULX-8, in addition to new observations, almost published all {\it Chandra} data sets (Table \ref{T:1}) were used. In order to obtain the long-term X-ray light-curve of ULX-8, the observed unabsorbed X-ray fluxes derived using 21 {\it Chandra} observations (Table \ref{T:1}) in the energy range of 0.5-7 keV. For this, a generic spectral shape was assumed with a {\it power-law} photon index of $\Gamma$=1.7 \citep{2020MNRAS.491.1260S} and a Galactic absorption component, N$_{H}$=0.03$\times$10$^{22}$ cm$^{-2}$ using {\it srcflux} task in {\scshape ciao}. For in input source and background regions, 8$\arcsec$ and 16$\arcsec$ radii were chosen, respectively. The minimum and maximum unabsorbed X-ray fluxes were derived as (1.40 $\pm$ 0.13)$\times$10$^{-13}$ and (10.30 $\pm$ 0.40)$\times$10$^{-13}$ erg cm$^{-2}$ s$^{-1}$. Based on distance of galaxy M51 (9 Mpc) these fluxes correspond to X-ray luminosites as (1.35 $\pm$ 0.13)$\times$10$^{39}$ and (10.95 $\pm$ 0.39)$\times$10$^{39}$ erg s$^{-1}$, respectively (see Table \ref{T:1}). The resulting X-ray light-curve of ULX-8 and its histogram are shown in the Figure \ref{F:xraylc}. As seen in this figure, the source reached its peak flux values in recently published observations. There was no strong evidence of the bi-modality and any periodic modulation in the {\it Chandra} X-ray light-curve.

\section{DISCUSSION \& CONCLUSION} \label{sec:3}

The main goal of this study, the precise optical positions of the neutron star M51 ULX-8 obtained via advanced astrometry based on the {\it Chandra} and {\it HST} observations as well as {\it GAIA} source catalogues. Examining all the optical images listed in Table \ref{T:foto} and also 2005-01-20 {\it HST}/ACS F435W, F555W and F814W data sets, three discrete sources C1 ,C2 and C3 were identified within the astrometric error radius of 0$\arcsec$.58 at 90\% confidence level. This result is different from the astrometric calculations given in \cite{2006ApJ...645..264T}. They found a unique optical counterpart that was not within the error radius derived in this study. In their study, all reference sources were not in the same frame on {\it HST} images. In addition, all references are not located on the optical axis of the S3 chip of {\it Chandra} ACIS. Moreover, there is also only 0$\arcsec$.05 offset between the {\it Chandra} X-ray and {\it HST} optical coordinates of the ULX-8 which corresponds to just one pixel in the ACS/WFC images. In this study, astrometric errors were calculated more precisely and different from their approach.

In order to search for any optical periodic modulations, 34 {\it HST} ACS/WFC F606W and F814W observations were used. A 125.5 $\pm$ 2.3 d optical periodic modulation was determined for counterpart C3 in the F606W and F814W long-term light-curves with amplitude of 0.14 $\pm$ 0.02 mag. at greater than 95\% confidence level. No periodic modulation and also no optical variability were observed for remaining counterparts C1 and C2.

The log(F$_{X}$/F$_{X}$) ratios for C1 and C3 were found as $\sim$ 2.6 and $\sim$ 2.3 respectively which are in the range of X-ray binaries values. However,it was found as $\sim$ 1.6 for C2 that this value is within the given range for AGNs or BL Lac objects. Therefore, the possibility that the C2 is the optical counterpart of the ULX-8 may be ruled out. However, considering the ULX inclination, we could not know whether a ULX can have F$_{X}$/F$_{optik}$ in the same range as an AGN. In addition, if these sources are truly the optical counterparts of the ULX-8 and also if the observed X-ray emission comes from the accretion flow, X-ray emission could change on far shorter timescales. Therefore, these ratios calculated from nearly simultaneous observations may be misleading for classifying sources. Since the C3 is almost located in the center of the 1-$\sigma$ (68\%) as well as 90\% astrometric error radius (see Figure \ref{F:can}) it could be the most likely optical counterpart of the ULX-8. Moreover, the remaining optical counterpart C1 is not completely excluded, but the optical source C3 has various evidence (e.g. X-ray-optical variability) to represent the optical nature of the ULX-8.\\

The ULX-8 which its compact object well-known to be as a strong neutron star candidate has shown long-term X-ray variability \citep{2020ApJ...895..127B}. However, there are no reported any period (e.g. spin, orbital or super-orbital) for the M51 ULX-8. The most exciting result of this study is the detection of an 125.5-d optical periodic modulation for possible optical counterpart C3. Since the orbital period of ULX-8 is unknown, this could be either an orbital period or a super-orbital period. The possible scenarios of the nature of optical periodic variation which are frequently mentioned in the literature are discussed for ULX-8 as follows:

\begin{itemize}

\item If a compact object interacted with a disk of a typical Be-X-ray binaries (BeXRBs) donor, the effects observed on the disk (e.g. darkening, rotating density waves) could be cause a super-orbital period \citep{2020MNRAS.495L.139T}. \cite{2020MNRAS.491.4949V} argued that the 38-day super-orbital period discovered for M51 ULX-7 was due to disk precession and they proposed free precession of the NS as a compelling motor to drive disc precession. The super-orbital periods seen in PULXs were correlated to the precession of an accretion disk induced by the Lense–Thirring precession of a large scale-height accretion flow \citep{2018MNRAS.475..154M,2019MNRAS.489..282M}. In most instances, super-orbital periods could be independently determined from optical light-curves of BeXRBs. When NS passes through around the equatorial disk of the rapidly rotating BeXRB donor, the orbital motion of the NS often causes a small optical flare. Also in this case, due to formation and depletion of disk or rotating density waves could be observed as a super-orbital period \citep{2020MNRAS.495L.139T}. The Be stars generally are massive main-sequence star and the range of orbital periods is (10–300) d with O-A types \citep{2009ApJ...707..870B}. If the 125.5 day is the orbital period of ULX-8, this scenario can be acceptable for C3. The super-orbital periods are also detected in the long-term X-ray light-curves of ULXs (e.g. NGC 5907 ULX1 \citealp{2016ApJ...827L..13W} and M51 ULX7 \citealp{2020ApJ...895..127B}) however, it was not found for the the ULX-8.\\

 \item The main problem with identifying the donor star in the optical emission of ULXs is that the optical contribution from the accretion disk, either directly or irradiated, is mostly unknown. In order to identify the possible contribution to the observed optical emission, the SED features of the counterparts were investigated using {\it HST} optical observations. In most cases, the observed {\it blackbody} spectrum of the optical emission is a feature of donor stars. On the other hand, if optical emission is mostly dominated by an accretion disk, the optical SED could have a roughly {\it power-law} spectrum as well as the optical-X-ray variability is also observed \citep{2011ApJ...737...81T}. These are quite consistent only in case C3 since the optical SED for only possible counterpart C3 is adequately fitted with {\it power-law} ($\alpha$$=$$-$ 2.98 $\pm$ 0.29) at 95\% confidence level (Figure \ref{F:SEDs}). Acceptable optical spectral models were not found for the C1 and C2.

 Due to the orbital motion of the donor which is distorted by the effects of gravity and is not irradiated equally by the X-ray radiation of the accretion disk, the optical periodic modulation that is not related to the orbital period could be observed. Although strong clues are required to accept the 125.5-day optical variability as a super-orbital period of disk origin, the possibility of this should not be ruled out. If indeed the periodic variability found for C3 is due to the orbital motion of the donor, the spectral type of C3 could not be predicted since the optical emission is dominated by the disk. Observed the X-ray-optical correlated variability could be a direct evidence of the irradiated disk model \citep{2011ApJ...737...81T,2012ApJ...745..123G}. In addition, optical variabilities of the Galactic X-ray binaries correlate with the X-ray variabilities \citep{2003astro.ph..8020C}. Simultaneous X-ray/optical observations are needed to constrain the nature of this variability. \\

 \item Double peaked profile (bi-modal distribution) of the light-curves may show evidences of precession of the disk or propeller effect \citep{2016MNRAS.457.1101T,2017AstL...43..464G}. However, the propeller regime is related to variabilities in the accretion rate and due to the fact that the observed flux modulation is periodic, it is incomprehensible how changes in the accretion rate can be periodic unless they are related to the binary orbit \citep{2019ApJ...873..115B}. \cite{2022MNRAS.510.4355A} reported an optical periodic variation of 264 d for one of the two optical counterparts of M51 ULX-4 which shows long-term X-ray variability two order magnitudes. In addition, they found some evidence for the existence of a bi-modality distribution of the X-ray flux of ULX-4. As seen in the Figure \ref{F:histo}, surprisingly, the observed optical fluxes of the C3 exhibits the bi-modal distribution in both filters which may suggest a link between high X-ray variability while there is no strong evidence of the bi-modality in the X-ray light-curve. Another optical periodic modulation (6.12-day) is discovered by \cite{2009ApJ...690L..39L} for the NGC 1313 X-2 which have weak coherent pulsations \citep{2019MNRAS.488L..35S}. This variability was explained as contamination from the X-ray reprocessing in the accretion disk \citep{2009ApJ...690L..39L,2012ApJ...745..123G,2012MNRAS.419.1331Z}. The scenario suggested for X-2 in which the shape of the optical light curve is similar to case C3 is quite reasonable for C3.\\

\item The CMD, deploying the V-band vs (B-V), was used to estimate the ages of the possible optical counterparts. The CMD of C1 and C3 was displayed in Figure \ref{F:CMD}. If the optical counterparts of ULX-8 are correlated with their environments, the age of C3 and C1 could be around 18-30 Myr and 90 Myr, respectively. The optical emission of ULXs may come from the accretion disk, either directly or irradiated. Therefore, these possibilities pose a problem for the identification of donor star of ULXs \citep{2022MNRAS.515.3632A}. If indeed, the optical emission mostly originate from the disk, the estimated age in case of C3 can be ignored.\\

\item Recently, studies (e.g. \citealp{2020ApJ...890..150C} and \citealp{2021ApJ...912...31H}) have suggested that the masses of the optical counterparts could be constrained depending on whether they are detected in the {\it HST} images. According to their studies, the possible counterparts of ULX-8 could be intermediate-mass X-ray binaries (IMXBs) or high-mass X-ray binaries (HMXBs) since considering the distance (9 Mpc) of galaxy M51, optical counterparts need to have a sufficiently high mass ($\geq$3 M$\odot$). On the other hand, it is also possible that the optical counterpart(s) of ULXs could be low mass X-ray binary (LMXB) and could not be detected in the astrometric error radius due to being intrinsically faint at the distance of galaxy M51. In addition, since many ULXs are located inside star forming regions (SFR). Since the SFR strongly dominates over the stellar mass, counterpart(s) might not be detected. Moreover, some ULXs are located in very dense gas and dust and also background level is very high for many galaxies therefore, even if ULXs have optical counterpart(s) they may not be detected.

\end{itemize}

\section*{Acknowledgements}
I would like to thank A. Akyuz for her valuable contributions and suggestions. The author gratefully thank to the Referee for the constructive comments and recommendations which help to improve the readability and quality of the paper. I would also like to thank my son Atlas ALLAK who is the sweetest source of my motivation.
\noindent

\section*{Data Availability}

The scientific results reported in this article are based on archival observations made by the {\it Chandra}\footnote{https://cda.harvard.edu/chaser/} and the NASA/ESA Hubble Space Telescope and obtained from the data archive at the Space Telescope Science Institute\footnote{https://mast.stsci.edu/portal/Mashup/Clients/Mast/Portal.html}

\bibliographystyle{mnras}
\bibliography{ulx8} 

\bsp	
\label{lastpage}
\end{document}